\documentclass[conference]{IEEEtran}
\IEEEoverridecommandlockouts
\usepackage{cite}
\usepackage{amsmath,amssymb,amsfonts}
\usepackage{algorithmic}
\usepackage{graphicx}
\usepackage{textcomp}
\usepackage{xcolor}
\usepackage{subfigure}
\usepackage{verbatim}
\usepackage{hyperref}
\usepackage{pifont}
\def\BibTeX{{\rm B\kern-.05em{\sc i\kern-.025em b}\kern-.08em
    T\kern-.1667em\lower.7ex\hbox{E}\kern-.125emX}}

\begin{document}

\title{Understanding the Challenges of Team-Based Live Streaming for First-person Shooter Games}

\author{\IEEEauthorblockN{Jiaye Li}
\IEEEauthorblockA{\textit{School of Science and Engineering} \\
\textit{ The Chinese University of Hong Kong, Shenzhen} \\
Shenzhen, China \\
jiayeli@link.cuhk.edu.cn}
\and
\IEEEauthorblockN{Minghao Li}
\IEEEauthorblockA{\textit{School of Science and Engineering} \\
\textit{The Chinese University of Hong Kong, Shenzhen} \\
Shenzhen, China \\
minghaoli1@link.cuhk.edu.cn}
\and
\IEEEauthorblockN{Zikai Alex Wen}
\IEEEauthorblockA{\textit{Computational Media and Arts Thrust} \\
\textit{The Hong Kong University of Science and Technology (Guangzhou)} \\
Guangzhou, China \\
zikaiwen@ust.hk}
\and
\IEEEauthorblockN{Wei Cai* \thanks{*Wei Cai is the corresponding author (caiwei@cuhk.edu.cn).}}
\IEEEauthorblockA{\textit{School of Science and Engineering} \\
\textit{The Chinese University of Hong Kong, Shenzhen} \\
Shenzhen, China \\
caiwei@cuhk.edu.cn}
}

\maketitle

\begin{abstract}
First-person shooter (FPS) game tournaments take place across the globe. A growing number of people choose to watch FPS games online instead of attending the game events in person. However, live streaming might miss critical highlight moments in the game, including kills and tactics. We identify how and why the live streaming team fails to capture highlight moments to reduce such live streaming mistakes. We named such mistakes \textit{jarring observations}. We conducted a field study of live streaming competitions of \textit{Game For Peace}, a popular FPS mobile game, to summarize five typical \textit{jarring observations} and identify three primary reasons that caused the issues. We further studied how to improve the live streaming system to prevent \textit{jarring observations} from happening by doing semi-structured interviews with two professional streaming teams for \textit{Game For Peace}. The study showed that a better system should (1) add a new sub-team role to share the director's responsibility of managing observers; (2) provide interfaces customized for three roles of live streamers in the team; (3) abstract more geographical info; (4) predict the priority of observation targets; and (5) provide non-verbal interfaces for sync-up between sub-teams. Our work provides insights for esports streaming system researchers and developers to improve the system for a smoother audience experience.
\end{abstract}


\section{Introduction}

Esports competitions have attracted widespread attention, with an increasing number of viewers opting to watch the competitions online~\cite{qian2020game, qian2020beyond}. Nonetheless, the audience often feels dissatisfied by missing game event highlights, such as missing kills and losing track of essential tactics. For example, hundreds of audiences complained about missing the kill moments in \textit{Counter-Strike: Global Offensive}~\footnote{https://blog.counter-strike.net/} Major that was streamed on \textit{Twitch}, the most famous esports streaming service~\cite{devia2017good}. On \textit{BiliBili}~\footnote{https://www.bilibili.com/}, a famous video-sharing website, people made videos to complain about the misplaced live streaming cameras that did not capture the highlight moments in \textit{League of Legends Pro League (LPL)}~\footnote{https://lpl.qq.com/}. Despite noticing so many complaints from the audience, there is insufficient research on understanding and addressing such complaints. Therefore, our work took one step further to understand how and why the live streaming teams make streaming mistakes, especially the mistakes they make due to the usability issues of the live streaming system.

The esports live streaming team generally consists of two sub-teams: the directors and the observers. The observers monitor the in-game data to keep track of ongoing events. They take charge of capturing highlight moments through a free camera. The directors select a video stream from all the observers' captured streams, which will be presented to the audience. If a team cannot cooperate smoothly for different reasons (e.g., talking over each other, operating the live streaming system wrong), then the audience will experience an acute sense of confusion and bewilderment, namely a \textit{jarring observation}. Although the live streaming teams can reduce the occurrences of \textit{jarring observations} by repeated practices, they cannot solve the problem because they need a reasonable team structure and a handy system to deal with the ever-changing game battlefield. Therefore, prior work proposed to introduce different new roles to the live streaming team~\cite{meixner2017multi} or to design algorithms for sports player identification and tracking~\cite{lu2013learning, baysal2015sentioscope,yang2021multi}. However, those proposals were designed for TV broadcasting on traditional sports, which does not help esports due to the distinct gaming nature~\cite{qian2020beyond}. 

Given that there are five major types of esports~\cite{funk2018esport}, we suspected that live streaming a different type of esports may require a different team-based work strategy and a different live streaming system. Therefore, we narrowed it down to focus on first-person shooter (FPS) games since they dominate the capital size of the esports tournament business~\cite{horst2021cs}. We conducted a field study on live streaming a competition of \textit{Game for Peace} to understand the challenges of team-based live streaming for FPS games. \textit{Game for Peace} is a famous battle royal shooter game~\footnote{https://en.wikipedia.org/wiki/Battle\_royale\_game/} that involves multiple players to compete under the last-man-standing gameplay mechanism, which is a branch of FPS games.

During the field study, we formalized the team's teamwork structure and identified the usability issues of the live streaming system used by the team. We found five typical \textit{jarring observations} including (1) the observers miss in-game firing; (2) The observers accidentally track the same player; (3) The observers ignore the director's instructions; (4) The observers switch to the wrong camera perspective of the player; And (5) the director's vague description that confuses the observers. 

We surveyed eight professional esports live streamers to study the potential reasons behind the cause of five \textit{jarring observations}. Based on the survey result, we summarized three primary reasons that cause five \textit{jarring observations}: (1) insufficient in-game information display, (2) lack of efficient communication channel between sub-team members, and (3) overlapping job duties. 

To derive suggestions for improving the teamwork structure and the design of a live streaming system, we conducted semi-structured interviews with two professional streaming teams for \textit{Game for Peace}. Based on the interviews, we proposed that an efficiently functioning live streaming team should consist of three live streaming roles that cooperate in a hierarchical structure: the director instructs the commander when the streams are not ideal; The commander monitors in-game events and assigns tasks to the observers; The observer monitors players and capture a video stream. In this case, a more useful live streaming system should (1) provide interfaces customized for three roles of live streamers in the team; (2) abstract more geographical information; (3) predict the observation priority of players automatically; and (4) provide non-verbal interfaces for sync-up between sub-teams. 

To summarize, our contributions are fourfold:
\begin{enumerate}
    \item We formalized the existing team-based streaming workflow for FPS games;
    \item We summarized the common interfaces design of a live streaming system for FPS games;
    \item We identified five typical \textit{jarring observations} and three primary reasons behind these issues;
    \item We proposed recommendations on reconstructing the teamwork structure and improving the design of a live streaming system.
\end{enumerate}

\section{Related Work}

There are few studies on reducing the occurrences of jarring observations during the live streaming of FPS games. Prior work focused on improving the traditional live sports TV streaming service through two approaches: optimizing the live streaming teamwork structure and designing algorithms for sports player identification and tracking.

Currently, live sports TV streaming service attempts to provide users the freedom to choose viewing resources, which may improve user engagement. However, the live streaming teams provide such service in an ad-hoc manner. They do not have a reasonable teamwork structure and a proper live streaming system. As a result, the audiences are often dissatisfied as they have bad experiences similar to \textit{jarring observations}. To improve the live sports TV streaming service, Britta et al.~\cite{meixner2017multi} conducted a field study and semi-structured interviews to formalize a reasonable teamwork structure and to identify the needs of the live streaming team, which shed light on designing a truly useful system for live sports TV streaming. 

Our work aims to address similar issues in FPS game live streaming: The live streaming teams work in an ad-hoc mode, and neither do they have a live streaming system designed for them. However, we cannot directly apply Britta et al.'s findings to help our target research object because FPS games have a distinct gaming nature. Therefore, we adapted Britta et al.'s research method to understand the issues in live streaming FPS games that cause \textit{jarring observations}.

Another approach to prevent \textit{jarring observations} is to design automated algorithms to assist the teamwork in judging which sports players will contribute to the highlight moments. However, traditional live sports TV streaming is still addressing the issue of identifying and keeping track of sports players in the video footage. The mainstream approach is to design deep learning algorithms to address the issue\cite{lu2013learning, baysal2015sentioscope, roggla2019lab, yang2021multi}. For example, Lu et al.~\cite{lu2013learning} and Yang et al.~\cite{yang2021multi} used different types of sports players' images to design supervised learning algorithms. However, few studies took a step further to utilize sports players' location and activity information results to predict the observation priority of players. Therefore, our work studied whether the AI prediction feature is needed by our target research object and how it should be integrated into the existing live streaming systems.

\section{Field Study and Survey}

Our first step is to study what kinds of \textit{jarring observations} may occur in practice and their possible patterns, so we conducted a field study in the real setting where the live streaming team worked for a two-day esports tournament of \textit{Game for Peace}, a battle royal shooter game created by \textit{Lightspeed Studio}. Then, we surveyed eight professional live streamers to find out the causes of \textit{jarring observations}.

\subsection{Methodology}

Based on the audience complaints of \textit{jarring observations}, we paid attention to identifying and categorizing the \textit{jarring observations} that happened during the live streaming service. During the field study, we also recorded how the live streaming team cooperates as a team. After the live streaming service ended, we prompted questions about how the teamwork structure formed and the functions of specific live streaming system interfaces. After the field study, we transcribed the communication information and used the open coding method~\cite{mills2009encyclopedia} to identify the common \textit{jarring observations}. 

Based on the results of the field study, we surveyed eight professional live streamers, including one director and seven observers, to study the causes of \textit{jarring observations}. To complete the survey, the participants first read the description of \textit{jarring observations} and then answered the following open-ended questions for each \textit{jarring observation} scenario:

\begin{itemize}
\item \textbf{Q1}: Did you encounter such kind of scenario? 
\item \textbf{Q2}: How does the scenario influence live streaming? What are the benefits and harms?
\item \textbf{Q3}: On a scale of one (definitely not) to five (definitely yes), how much does the scenario negatively affect a live streaming experience? 
\item \textbf{Q4}: How would you deal with the scenario?
\end{itemize}

We used a mixed-method approach to analyze the result of the surveys.

\subsection{Teamwork Structure}

In general, the live streaming team for FPS esports is divided into two sub-teams: the director and the observer(s). The director instructs observers to monitor specific players and capture the players' highlight moments. The director can receive multiple video streams from the observers on a big monitor screen. Occasionally, the observers may speak up loudly to update an ongoing event to the whole team. 

\begin{figure}[htbp]
\centering
\includegraphics[width = 0.7\columnwidth]{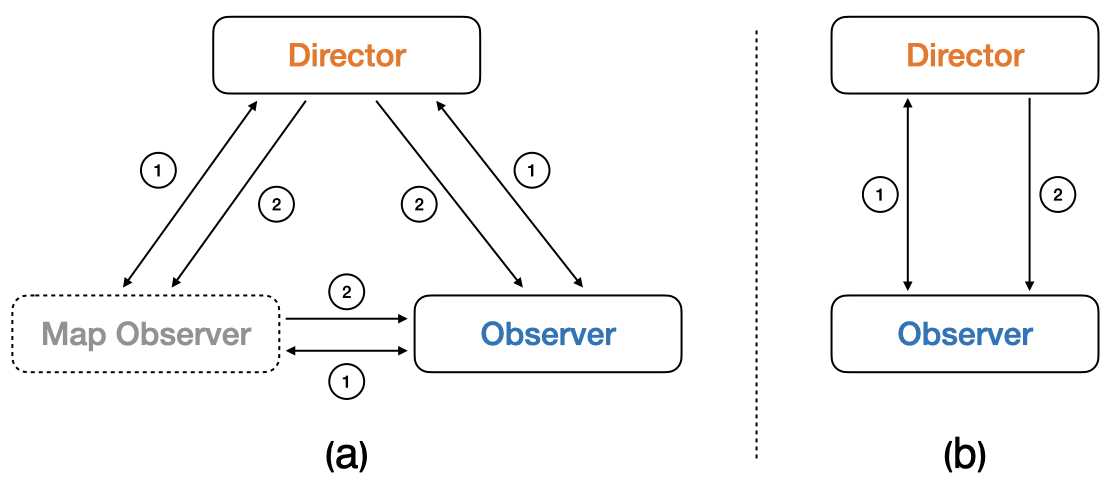}
\caption{Potential teamwork structures and their information flow. (a) illustrates the teamwork structure that we observed in the field study. (b) illustrates the general teamwork structure. \ding{192} indicates in-game data communications and sync-up. \ding{193} indicates the instructions to update the video stream. }
\label{Img3}
\end{figure}

In our field study, we further observed that the live streaming team refined their teamwork structure by delegating one of the observers only to take charge of reporting the critical events observed from the game's mini-map, which is called a map observer. The team's overall structure and information flow are shown in \autoref{Img3}. The map observer shares the director's responsibility to monitor critical in-game events. However, we found from our field study that this position did not yet address how players communicate with each other by shouting out the messages, which caused chaotic moments that led to \textit{jarring observations}.

\subsection{Live Streaming System Interfaces}

\begin{figure}[h!]
\centering
\subfigure[First-Person View]
{
    \includegraphics[width = 0.45\columnwidth]{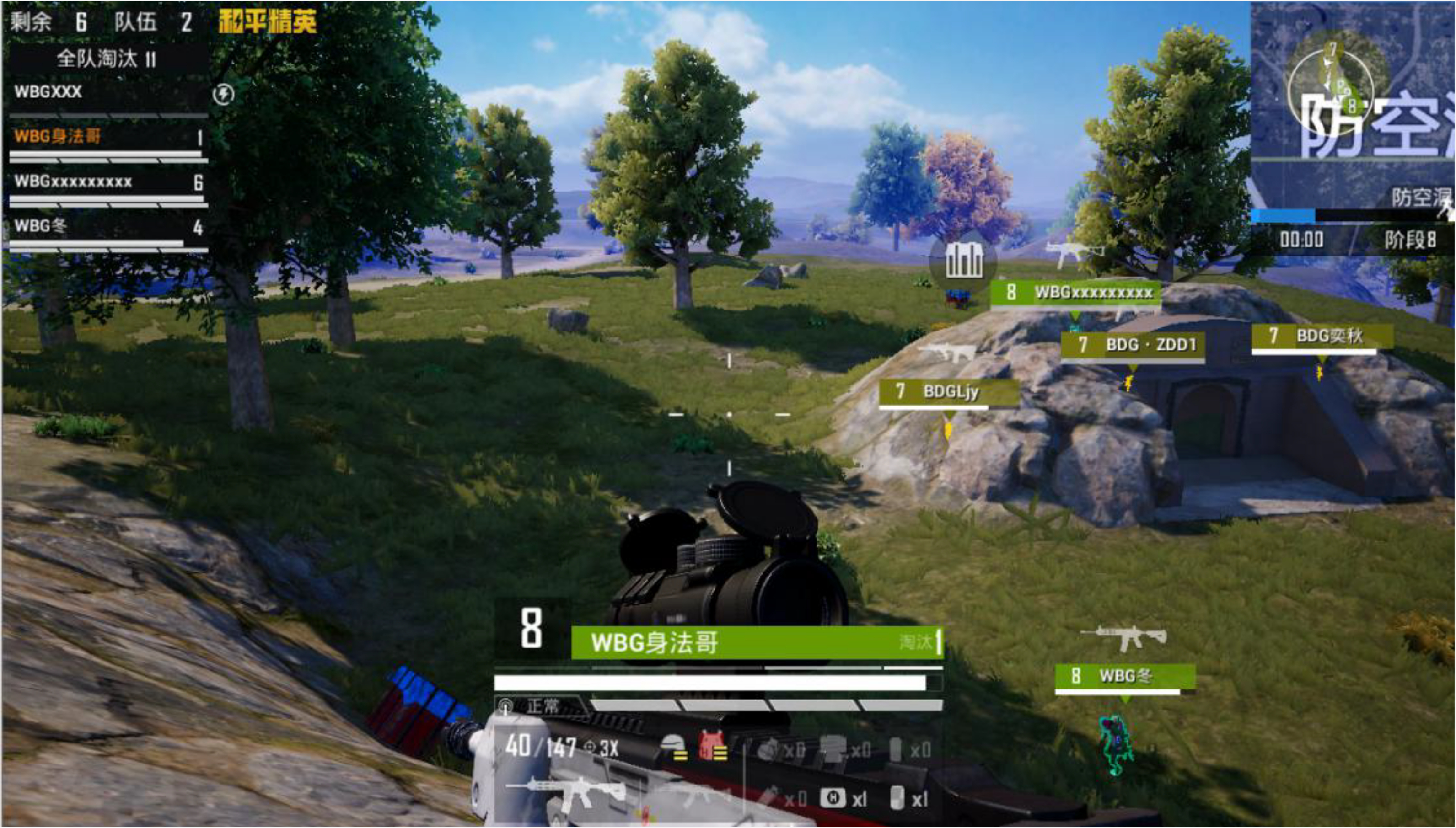}
}
\subfigure[Third-Person View]
{
    \includegraphics[width = 0.45\columnwidth]{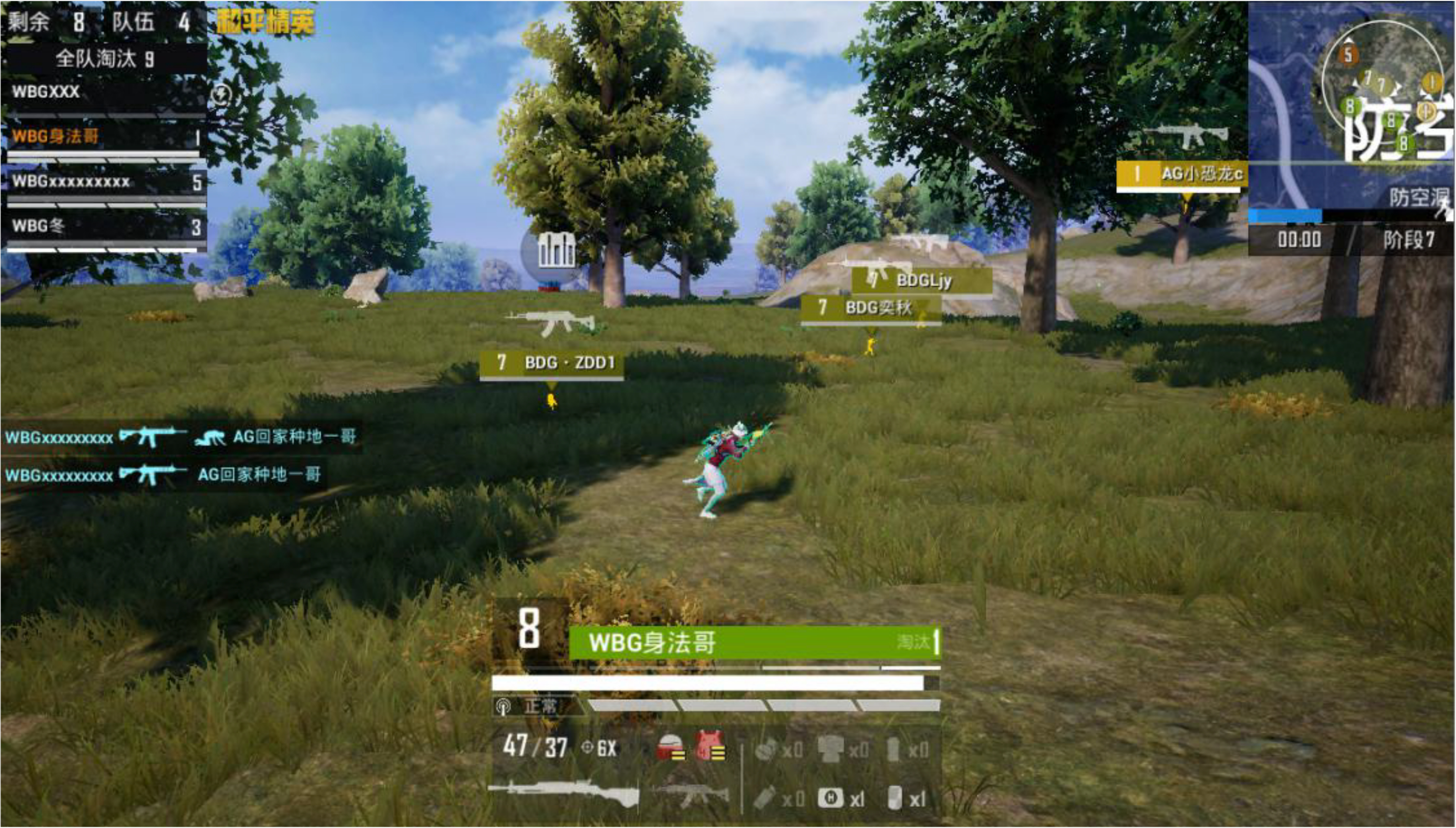}
}

\subfigure[Free-Roam View]
{
    \includegraphics[width = 0.45\columnwidth]{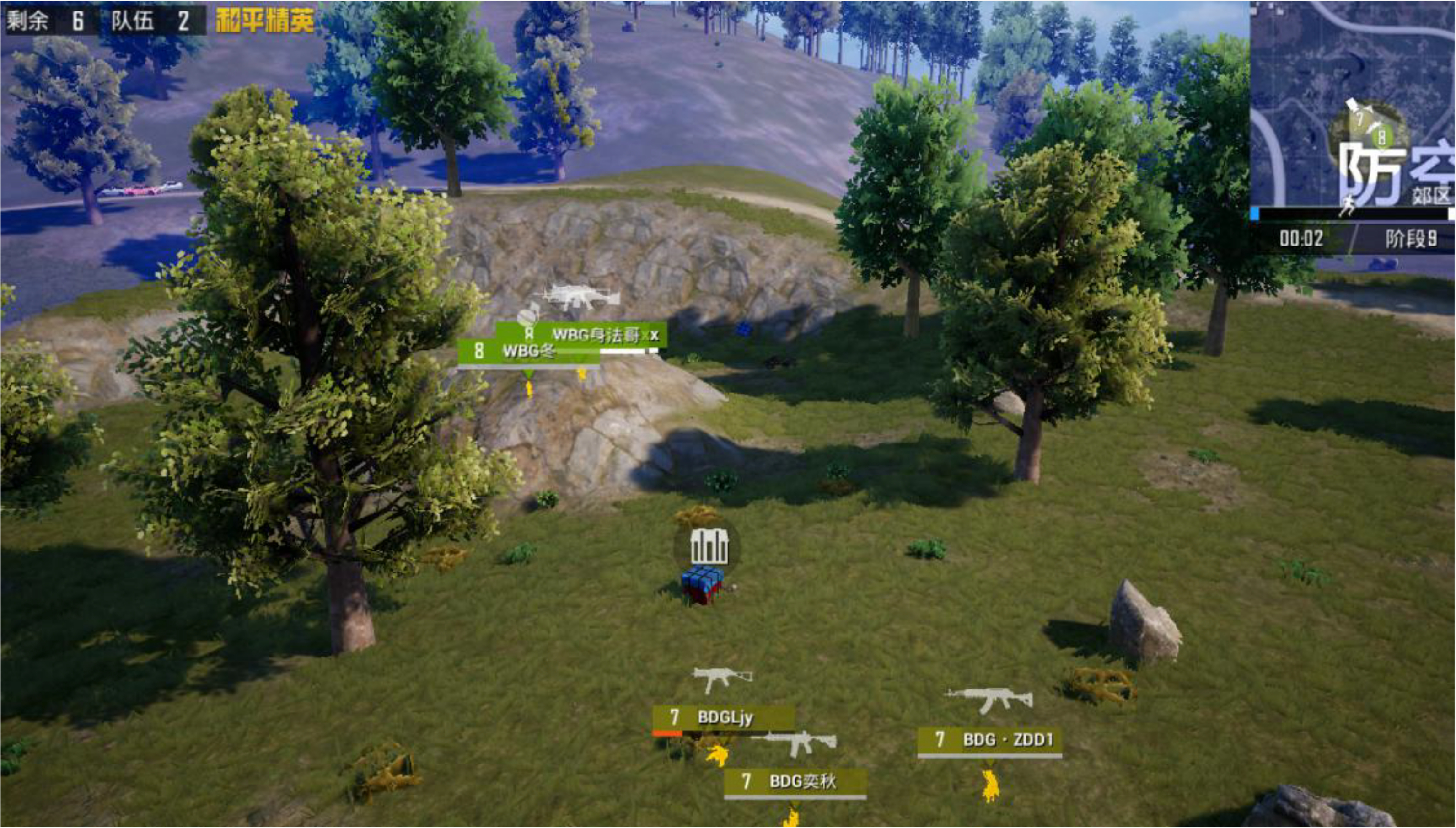}
}
\subfigure[Follow Mode View]
{
    \includegraphics[width = 0.45\columnwidth]{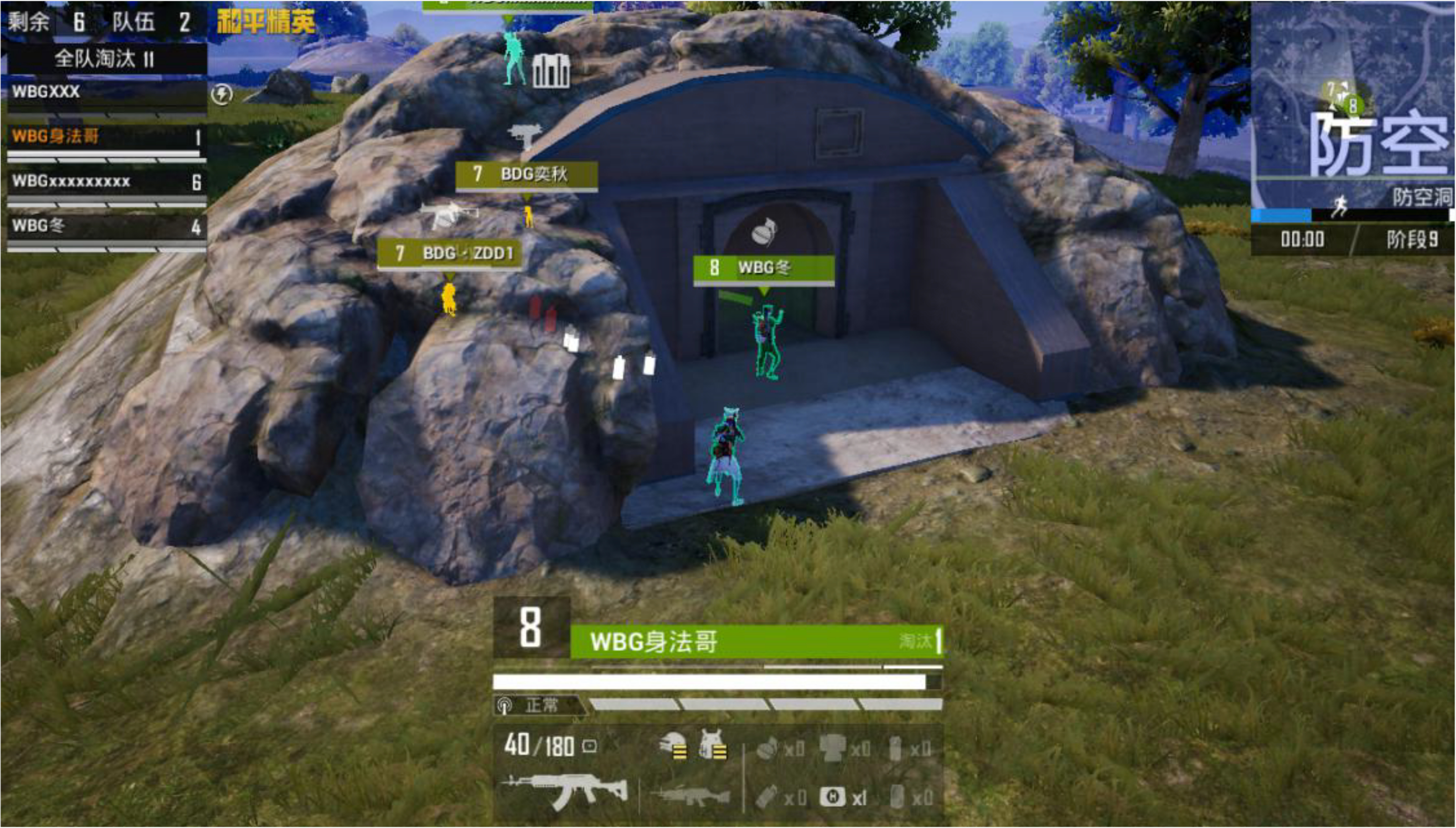}
}
\caption{The examples of four camera views that the observer can operate.}
\label{Img4}
\end{figure}

In our field study, we found that the live streaming system provided different interfaces for the director and the observers. The observers mainly operated the virtual in-game camera in four modes as shown in~\autoref{Img4}: a first-person view, a third-person view, a free-roam view, and a follow mode view. We did not find that this camera operations design might potentially cause \textit{jarring observations}.

\begin{figure}[htbp]
\centering
\subfigure[``Wall Hack'']
{
        \includegraphics[width = 0.45\columnwidth]{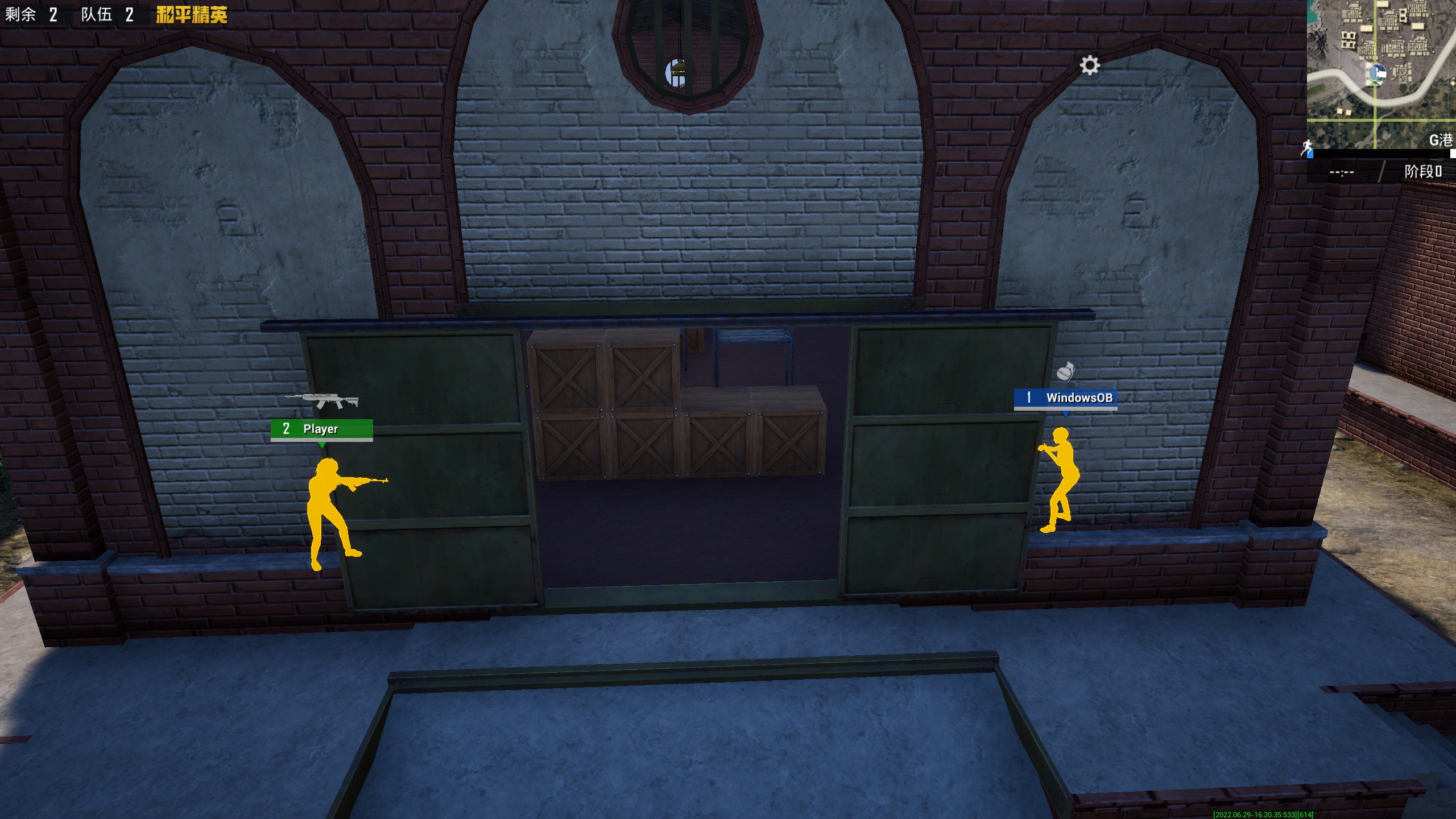}
}
\subfigure[Utility Trajectory]
{
        \includegraphics[width = 0.45\columnwidth]{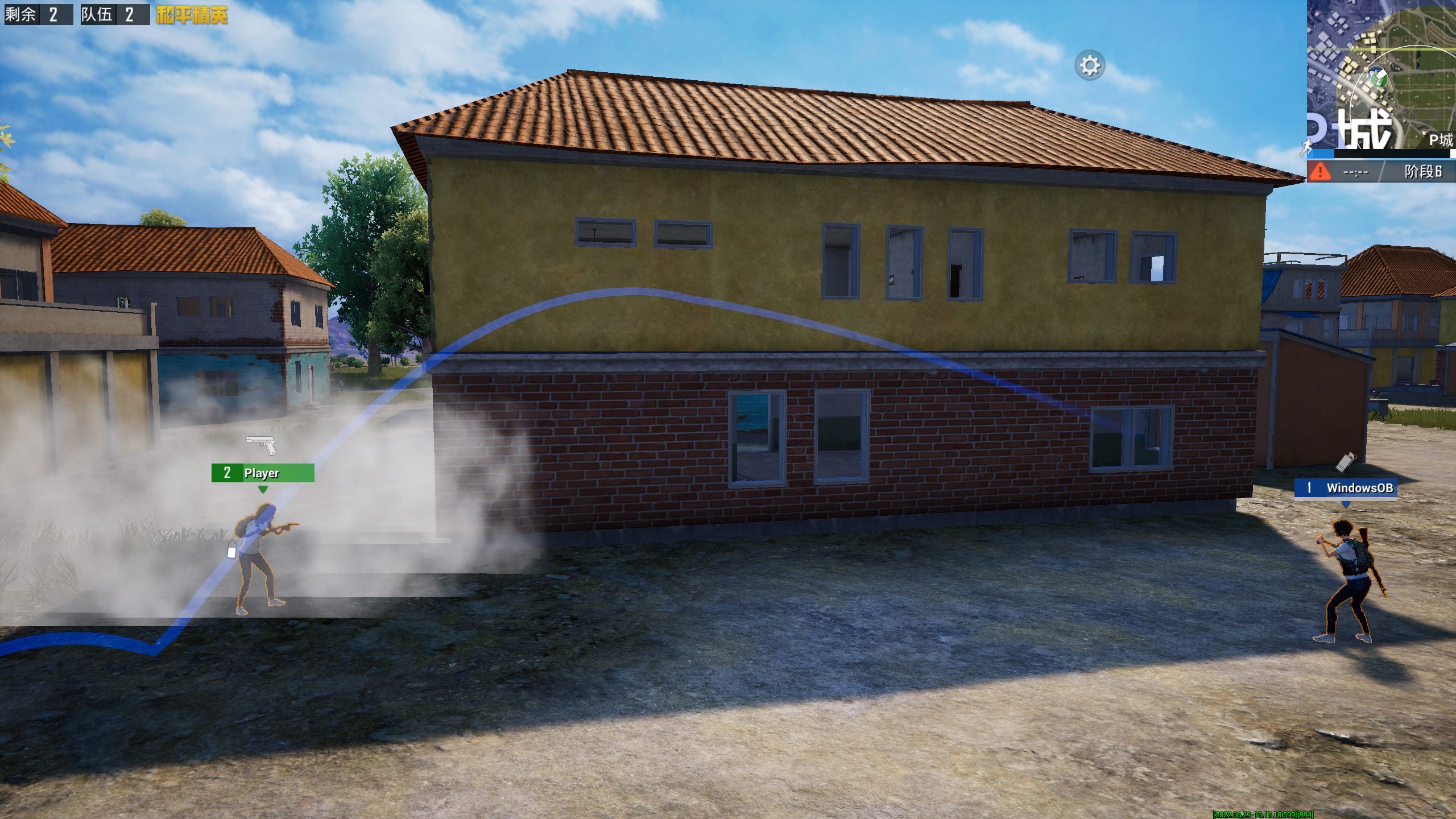}
}
\caption{The examples of the ``Wall Hack'' function and the utility trajectory visualization on the observer's screen.}
\label{ImgWHUT}
\end{figure}

The observers also showed their system's front-end display that provides the following information: players' names, equipment, utility trajectory, and kills. To keep tracking the player hidden behind a wall, the observer can enter the ``Wall Hack" mode to see through a physical wall. The examples of the ``Wall Hack" mode and the utility trajectory visualization are shown in \autoref{ImgWHUT}. Neither did we find that the design of the front-end display might potentially cause \textit{jarring observations}.

\begin{figure}[htbp]
\centering
\centerline{\includegraphics[width = 0.7\columnwidth]{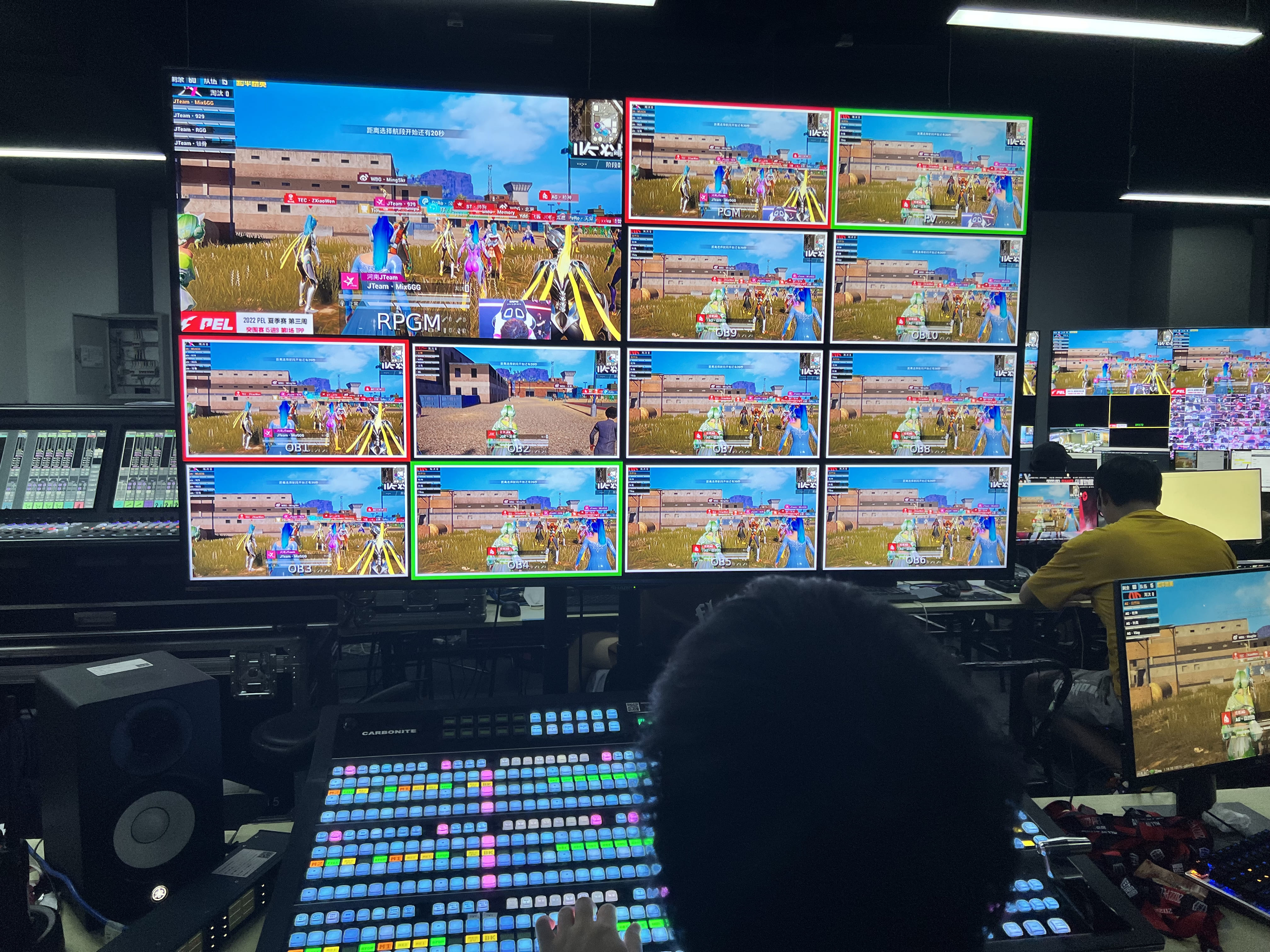}}
\caption{The multi-screen view for the director. The largest screen on the left top of the screen is the current streaming video. Two screens on the right top are the PGM and PV. The rest of the screens receive the live video signals from the observers. }
\label{DirectorView}
\end{figure}

The director sits in front of a huge multi-screen monitor as shown in~\autoref{DirectorView}. The multi-screen monitor receives live video signals coming from the observers. The monitor provides two additional special screens besides the observers' screens: The PGM screen displays the current video signals that the audience is watching; The PV screen displays the video signals from another source that is ready to switch to. The director assigned an observer's streaming video to one of the special screens through a physical button on a keypad. We found that the director was distracted by not being able to maintain an intensively high level of focus while managing multiple screens for a long period, which resulted in some \textit{jarring observations}.

\subsection{Typical Jarring Observations and Primary Causes}

We summarized five typical \textit{jarring observations} that we recorded in the field study:

\begin{itemize}
\item \textbf{S1}: The director found that a player was under attack and decided to switch to the attacker's first-person (FP) perspective. However, none of the observers provided such a camera view. Thus, the observers had to search for the attacker after the director shouted out this issue;
\item \textbf{S2}: Because the observers only knew their camera status, they were unclear which observer's video source was being used by the director, and also, the observers accidentally tracked the same player;
\item \textbf{S3}: Multiple fights among players broke out during a short period. The director's instruction was interrupted by fierce communication among all team members, so the observers ignored the instruction; 
\item \textbf{S4}: The observer whose video signal was streaming on PGM was supposed to direct the camera at the shooter but accidentally directed it to the player being shot;
\item \textbf{S5}: The director used ``that yellow jacket" to describe a player, but the observers insisted on not finding any player wearing a yellow jacket.
\end{itemize}

\begin{figure}[htbp]
\centering
\subfigure[]
{
        \includegraphics[width = 0.45\columnwidth]{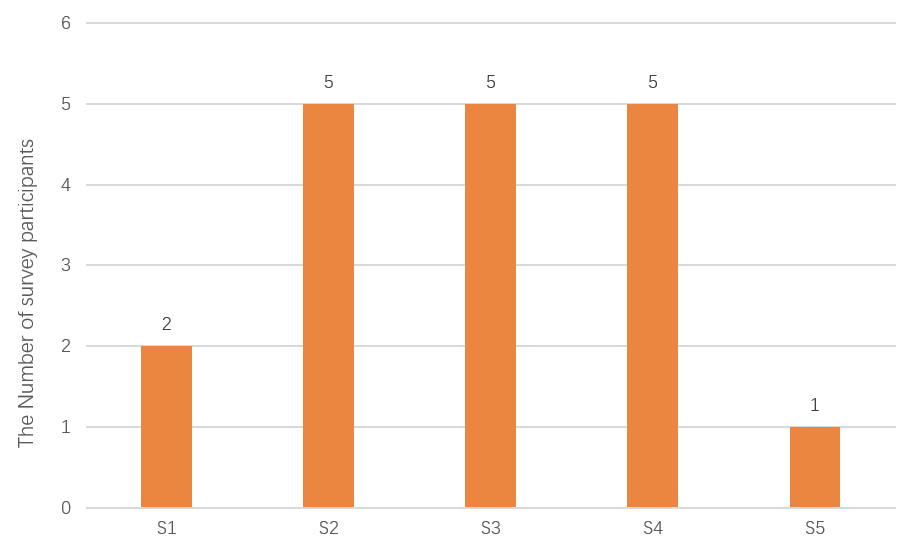}
}
\subfigure[]
{
        \includegraphics[width = 0.45\columnwidth]{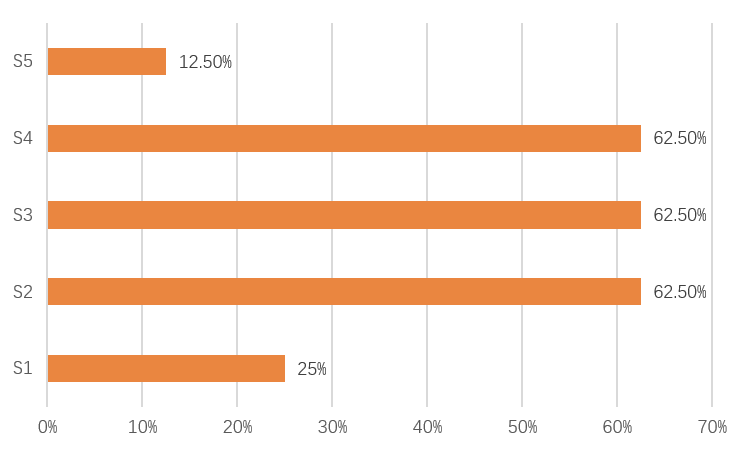}
}
\caption{(a) is a bar chart that shows the number of participants who experienced the \textit{jarring observation} scenario. (b) is a bar chart that shows the percentage rate of the bar chart (a). The top percentage of experienced scenarios are \textbf{S2}, \textbf{S3} and \textbf{S4}; and their percentages are 62.5\%. 25\% of participants experienced \textbf{S1} and 12.5\% for \textbf{S5}.}
\label{Number}
\end{figure}

The results of the survey \textbf{Q1} showed that some of the \textit{jarring observations} are commonly experienced, but some are not. As shown in~\autoref{Number}, five out of eight live streamers mentioned that they had jarring observations in scenarios \textbf{S2}, \textbf{S3}, and \textbf{S4}, which seems to be common. Meanwhile, the live streamers did not commonly experience the scenarios \textbf{S1} and \textbf{S5}.

\begin{figure}[htbp]
\centering
\centerline{\includegraphics[width = 0.8\columnwidth]{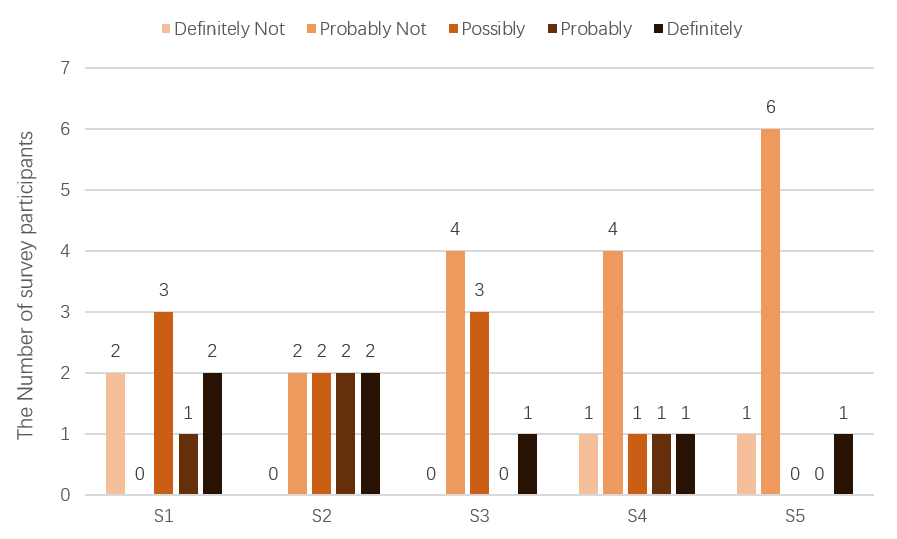}}
\caption{The diagram that shows the number of participants rating the level of each scenario's negative impact. The total number of participants is eight.}
\label{ImpactLevel}
\end{figure}

Our survey participants also mentioned that the five \textit{jarring observation} scenarios harm the live streaming experience. As shown in \autoref{ImpactLevel}, the scenarios \textbf{S1}, \textbf{S2}, and \textbf{S3} have a highly negative impact on the live streaming experience, while \textbf{S4} and \textbf{S5} have a moderate negative effect. 

Based on the survey result of \textbf{Q4}, we derived the underlying causes for each scenario: 

\begin{itemize}
\item \textbf{C1 $\Rightarrow$ S1}: The observers have to operate the virtual camera using a keyboard or a mouse to allocate the target player's position. So, the observers are too busy searching for the target on the map to miss highlight moments; 
\item \textbf{C2 $\Rightarrow$ S2}: The observers could not synchronize which player to follow because they did not have the information about other observers' screens and the source of PGM and PV signals;
\item \textbf{C3 $\Rightarrow$ S3}: The observers shouted out loud to sync up on live streaming tasks so that they may miss the director's instructions because of the noisy working environment;
\item \textbf{C4 $\Rightarrow$ S4}: Because the kills per minute are extremely high in FPS games, so the observers find it difficult to predict which players may be the next killer, which was quoted as ``it is hard to predict who would make the first shot'';
\item \textbf{C5 $\Rightarrow$ S5}: The observers find it hard to understand vague descriptions from the director (e.g., the ``that yellow jacket guy'' description).
\end{itemize}

In summary, we found from \textbf{C1 $\Rightarrow$ S1} and \textbf{C4 $\Rightarrow$ S4} that the design of a live streaming system should provide at least two more front-end features for the observer to track down their target player: locomotion to quickly allocate a player's position and AI prediction about which player might be the next most likely killer. We also found from \textbf{C3 $\Rightarrow$ S3} and \textbf{C5 $\Rightarrow$ S5} that the existing live streaming systems do not provide helpful communication channels for team members to communicate across the sub-teams. \textbf{C2 $\Rightarrow$ S2} pointed to the issue that the teamwork structure needs to be refined in order to reduce overlapping job duties, which may reduce the occurrences of \textit{jarring observations}. Live streaming teams are more likely to keep making \textit{jarring observation} mistakes if they still have to rely on voice communication~\cite{raniadvantages}.

\section{Interview Study with Live Streaming Teams}

In order to propose useful design recommendations to prevent the \textit{jarring observations} from happening, we conducted two interviews with two professional FPS games live streaming teams. The interviews were conducted on \textit{Tencent Meeting}, an online meeting tool, and lasted between 90 and 120 minutes. 

\subsection{Methodology}

Before the interview, the research team met with each participant individually to explain the goal of the interview, answer any questions, and understand any accessibility needs. Digital materials were shared with all participants one week in advance, including background information and a detailed description of the five \textit{jarring observations}. 

The first interview focused on the brainstorming stage. The goal of this interview was to encourage interviewees to provide opinions on massive data management and communication efficiency improvement. The interview began with the facilitators briefly introducing the background information of this research, including the five typical \textit{jarring observation} scenarios. And then, participants were encouraged to share unpleasant \textit{jarring observation} experiences during team streaming. By communicating with participants, the facilitators located more elements that affect streaming performance. The facilitators then prompted questions for participants to reflect on their responses to avoid \textit{jarring observations}. The interview concluded with a brainstorming activity where facilitators prompted participants to propose more suggestions on \textit{jarring observation} prevention. Throughout the discussions, the facilitators took notes to record the participants' feedback. 

Between the first and second interviews, the research team formed a collective summary of suggestions based on the first interview's conclusion. The facilitators provided a summary of what was learned from the interview and answered any of the participants' questions that could not be explained during the interview through \textit{WeChat}. The second interview focused on prompting participants to polish suggestions from the first interview. The interview again concluded with a summary of the main insight learned. 

After the interview, we transcribed observation notes and interview recordings from both interviews. We used open coding to organize the data and identify common findings. 

Six participants were from live streaming teams for the \textit{Game for Peace} esports. All participants are professional live steamers in China, with a specialty in the FPS games. The median working seniority is 2.5 years (SD = 0.84, range = 2.5) and the median age was 25.5 (SD = 0.75, range = 0.75). Two participants serve as Directors in the team streaming, while the remaining four are observers. We include three participants (one director and two observers) in each interview round. Additionally, two research team members participated as facilitators during both interviews. Both facilitators have hundreds of hours of FPS game experience. 

\subsection{Design Recommendations to Prevent Jarring Observations}
Findings from the interview suggest improving team structure of the live streaming team, enriching game data display, and providing AI prediction features during live streaming. We conclude with five design recommendations to help the team prevent \textit{jarring observations}:

\begin{itemize}
    \item \textbf{RD1}: Add a new sub-team role, the commander, to share the director's responsibility of managing observers;
    \item \textbf{RD2}: Provide interfaces customized for three roles of live streamers in the team;
    \item \textbf{RD3}: Abstract more geographical information;
    \item \textbf{RD4}: Predict the observation priority of players;
    \item \textbf{RD5}: Design non-verbal interfaces for sync-up between sub-teams.
\end{itemize}

We explain how our findings support these suggestions in the following paragraphs. 

\begin{figure}[htbp]
\centering
\centerline{\includegraphics[width = 0.7\columnwidth]{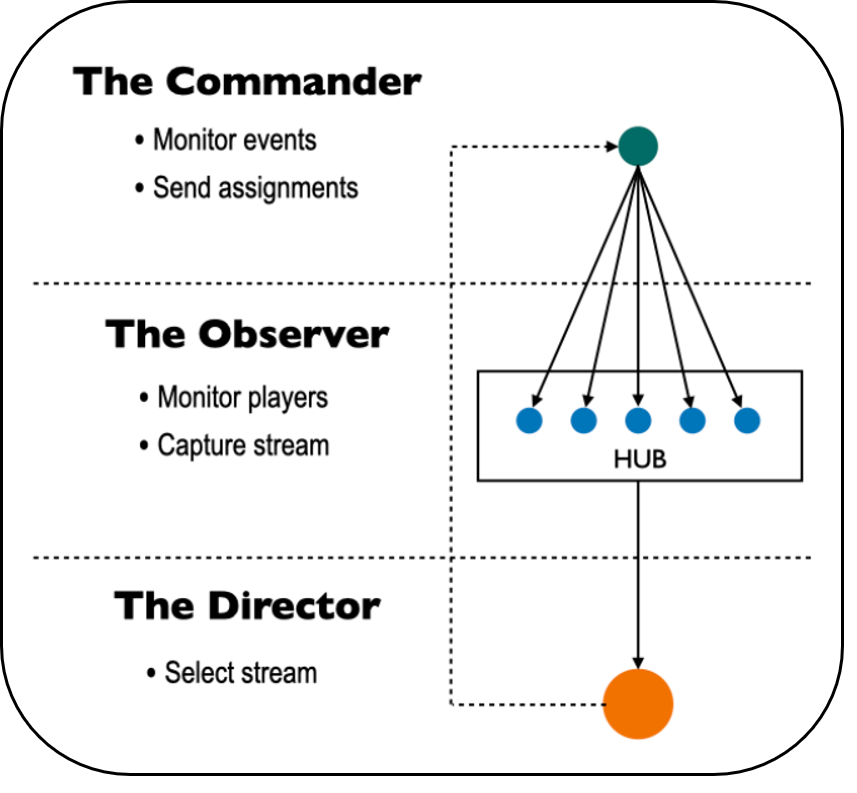}}
\caption{A novel team structure instance for live streaming. The commander monitors events and assigns tasks to observers. The observers stream their target players' activities based on the commander's instructions. Observers' streaming signals are live on the director's monitor screen. The director gives instructions to the commander if there is any missing camera view. }
\label{Findings}
\end{figure}

Team structure in the current live streaming team is redundant since each position's responsibility is unclear, which causes overlapping workflow. It is necessary to add a new position between the director and the observer, whose duty includes in-game events sync-up and issuing tracking instructions to observers. Moreover, the director should only take charge of selecting scenes for streaming, while the observer should only focus on capturing highlights. As a director suggested, ``I think we should not provide too much information for the director. Since I need to keep an eye on around ten screens, any other irrelevant information distracts for me.'' In actual practice, some professional live streaming teams set a subgroup named ``Scheduler'' which ``tracks all game teams' moving tendencies, updates battle situations, and predicts the next potential highlights''. By adding this subgroup and simplifying the duties of the Director and the Observer, we can reach ''a top-down workflow'' and ``alleviate communication stress during live streaming''. Based on our interviews, we proposed a novel team structure in \autoref{Findings}. (\textbf{RD1}, \textbf{RD2})

The participants were passionate about displaying more in-game data during live streaming. Current significant yet unrevealed data includes detailed distance between players and distinguishable map terrain information. The participants illustrate the importance of distance and terrain with the example, ``When two teams move close to each other, there is a high possibility for combat except one of them on a cliff, and the other down below, which will probably be no fight happen.'' The unclear presentation of terrain in the current system, especially indistinct altitude information, makes it difficult for observers to evaluate whether a combat would happen. The current mini map ``already has color distinguished altitude display.'' However, ``players'(represented by colored points in the mini-map) color may cover the map's color,'' which causes the observers unable to distinguish height. (\textbf{RD3})

\begin{figure}[htbp]
\centering
\centerline{\includegraphics[width = 0.5\columnwidth]{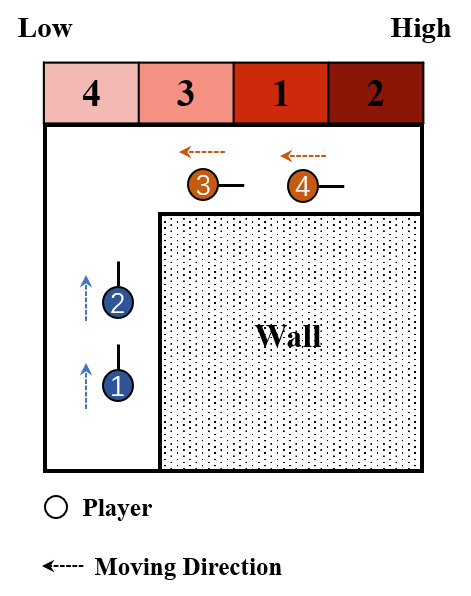}}
\caption{This is an example of how to define observation priority. Four players from two teams move towards a corner with P1, P2 facing the corner, and P3, P4 back towards the corner. Under this circumstance, P2 is most likely to shoot the enemy first; therefore, P2 is a the top of the observation priority. With P1 following P2, P1 would be the second on the list. P3 and P4 will be the third and the final. }
\label{ObservationTier}
\end{figure}

Another vital element for FPS live streaming is the \textit{observation priority} of players. \autoref{ObservationTier} is an example of how to define \textit{observation priority}. Observation priority is usually formulated at the beginning of the game, and updates in synchronized with current game events. Currently, the live streaming team must report the \textit{observation priority} by shouting out aloud or relying on team members' experiences. Participants suggested that the \textit{observation priority} should be updated to each team member without latency. In addition, researchers have developed AI algorithms~\cite{lu2013learning, baysal2015sentioscope,yang2021multi} to identify and track sports players in recent years, which showed excellent quality. Therefore, we advocate that the research community may explore AI algorithms to assist the live streaming teams in predicting the \textit{observation priority} of players. (\textbf{RD4})

In practice, team members depends on voice communication to sync-up. Multiple simultaneous in-game events may cause conflict or overlapping voice communication among teammates, which results in miscommunication. The participants need to identify teammates by their voices and pick tasks, which requires experienced live streaming skills and a massive amount of training. Our interview participants strongly suggested adding non-verbal interfaces for sync-up. One participant said, ``it would be very helpful if I could mark on the mini-map for others to see.'' (\textbf{RD5})

\section{Conclusion and Future Work}
\textit{Jarring observation} is the issue of missing highlight moments in live streaming an esports event. Our work analyzed the existing \textit{jarring observations} in FPS esports live streaming and studied the causes of creating \textit{jarring observations}. We then conducted two interviews with professional teams to propose design recommendations to help the live streaming team prevent \textit{jarring observations}. Our interviews focused on improving the teamwork structure, enriching in-game data display, predicting the observation priority of players, and adding non-verbal interfaces to facilitate sync-up between sub-teams. In the future, we plan to design a system prototype based on our findings and apply our recommendations to real FPS esports live streaming to test our system design.
    
\section*{Acknowledgment}

This research was supported by \textit{LightSpeed Studios}. The authors would like to thank \textit{LightSpeed Studios} for coordinating the field study and arranging participants for the interview study. We would also like to thank our study participants and anonymous reviewers for their valuable inputs. 

\bibliographystyle{plain}
\bibliography{ref.bib}
\end{document}